\documentclass[prapplied,aps,superscriptaddress,showpacs,reprint]{revtex4-2}
\usepackage{graphicx}  
\usepackage{dcolumn}   
\usepackage{amssymb}
\usepackage{physics}
\usepackage{color}
\usepackage[dvipsnames]{xcolor}
\usepackage[colorlinks=true, citecolor={blue!80!black}, urlcolor={blue!50!black}, linkcolor = {blue!80!black}]{hyperref}
\DeclareUnicodeCharacter{2212}{-}

\begin{document}

\title{Magnetic-Field-Compatible Superconducting Transmon Qubit}

\author{A.~Kringh{\o}j}
\thanks{These authors contributed equally.}
\author{T.~W.~Larsen}
\thanks{These authors contributed equally.}
\author{O.~Erlandsson}
\affiliation{Center for Quantum Devices, Niels Bohr Institute,
University of Copenhagen, 2100 Copenhagen, Denmark}
\affiliation{Microsoft Quantum Lab-Copenhagen, Niels Bohr Institute,
University of Copenhagen, 2100 Copenhagen, Denmark}
\author{W.~Uilhoorn}
\affiliation{QuTech and Kavli Institute of Nanoscience, Delft University of Technology, 
2600 GA Delft, The Netherlands}
\author{J.~G.~Kroll}
\affiliation{QuTech and Kavli Institute of Nanoscience, Delft University of Technology, 
2600 GA Delft, The Netherlands}
\author{M.~Hesselberg}
\author{R.~P.~G.~McNeil}
\affiliation{Center for Quantum Devices, Niels Bohr Institute,
University of Copenhagen, 2100 Copenhagen, Denmark}
\affiliation{Microsoft Quantum Lab-Copenhagen, Niels Bohr Institute,
University of Copenhagen, 2100 Copenhagen, Denmark}
\author{P.~Krogstrup}
\affiliation{Center for Quantum Devices, Niels Bohr Institute,
University of Copenhagen, 2100 Copenhagen, Denmark}
\affiliation{Microsoft Quantum Materials Lab-Copenhagen, 2800 Lyngby, Denmark}
\author{L.~Casparis}
\affiliation{Center for Quantum Devices, Niels Bohr Institute,
University of Copenhagen, 2100 Copenhagen, Denmark}
\affiliation{Microsoft Quantum Lab-Copenhagen, Niels Bohr Institute,
University of Copenhagen, 2100 Copenhagen, Denmark}
\author{C.~M.~Marcus}
\author{K.~D.~Petersson}
\affiliation{Center for Quantum Devices, Niels Bohr Institute,
University of Copenhagen, 2100 Copenhagen, Denmark}
\affiliation{Microsoft Quantum Lab-Copenhagen, Niels Bohr Institute,
University of Copenhagen, 2100 Copenhagen, Denmark}

\begin{abstract}
We present a hybrid semiconductor-based superconducting qubit device which remains coherent at magnetic fields up to 1~T. The qubit transition frequency exhibits periodic oscillations with magnetic field, consistent with interference effects due to the magnetic flux threading the cross section of the proximitized semiconductor nanowire junction. As induced superconductivity revives, additional coherent modes emerge at high magnetic fields, which we attribute to the interaction of the qubit and low-energy Andreev states.

\end{abstract}
\maketitle

\section{Introduction}
Superconductor-semiconductor-superconductor (S-Sm\\ -S) nanowire Josephson junctions have been integrated into various superconducting circuits, including gate voltage tunable transmon qubits, known as gatemons~\cite{deLange:2015gv, Larsen:2015cp}, tunable superconducting resonators~\cite{Casparis:2018ac}, and Andreev qubits~\cite{Janvier:2015uf, Hays:2017ud}. These hybrid junction elements allow \textit{in situ} voltage control of their Andreev spectra and current-phase relation~\cite{doh_2005, Goffman:2017ab, vanWoerkom:2017ek, Spanton:2017ab}, in turn influencing measurable qubit properties such as anharmonicity~\cite{Kringhoj:2018ul} and charge dispersion~\cite{arno, dispersion}. Moreover, S-Sm nanowires in the presence of strong magnetic fields may host Majorana zero modes - as evidenced by both dc tunneling and Coulomb blockade spectroscopy measurements~\cite{Mourik:2012je, Albrecht:2016cw} - potentially forming the basis of robust topological qubits~\cite{Lutchyn2018}.

Recent work has demonstrated the coherent operation of gatemons with S-Sm-S nanowire junctions at moderate magnetic fields, $\sim100$ mT~\cite{Luthi:2018aa, sabonis2020}. Spectroscopy of S-Sm-S nanowire fluxonium qubits~\cite{pita2020} and graphene-based gatemons~\cite{Kroll:2018sd} at high magnetic fields ($\sim1$ T) has also been shown. However the detailed spectrum and time-domain coherence properties of gatemons at large magnetic fields remain unexplored. Realizing a magnetic-field-compatible transmon qubit would open a number of possible new research directions. For instance, a direct Josephson coupling of Majorana zero modes on separate topological superconductors is expected to modify the energy spectrum of a transmon qubit~\cite{Ginossar:2014aa,Yavilberg:2015aa}, offering a potential route to time domain studies of topological systems. Studies of other sub-gap features such as Andreev bound states~\cite{Hays:2017ud, Tosi_2019} could also take advantage of similar magnetic-field-compatible microwave circuitry. Furthermore, transmon qubits that can operate in high magnetic fields might enable control of a variety of spin ensemble-based quantum memories~\cite{Imamo:2008aa, ranjan_2013} or allow the origin of $1/f$ flux noise to be further elucidated through studying the polarization of spin impurities~\cite{oliver_2013, Luthi:2018aa, Kumar_2016}.

\begin{figure}
\centering\vspace{-0mm}
\includegraphics[width=1\columnwidth]{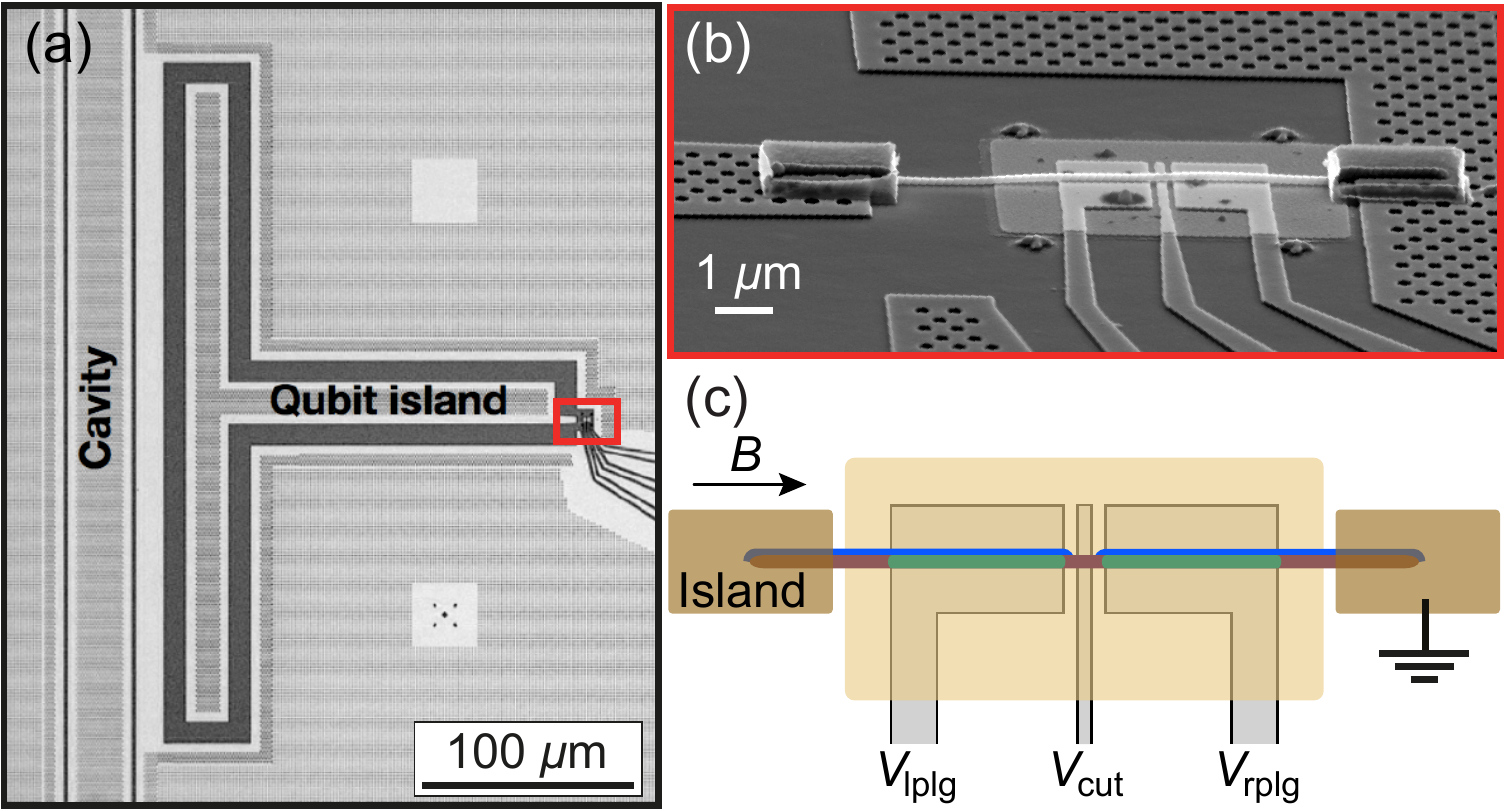}
\caption[Sample schematic]{\label{fig:SampleSchematics} Magnetic field-compatible device.
(a) Micrograph of the transmon qubit island capacitively coupled to a $\lambda/2$ cavity for readout and microwave control. The ground planes and inner conductors of the island and resonator were patterned with a large density of flux pinning holes for compatibility with large magnetic fields. A nanowire was placed to the right of the island (red rectangle). (b) Scanning electron micrograph of nanowire and bottom gates. One side of the wire junction was connected to the island and the other side was connected to the ground plane. (c) Device schematic showing an InAs nanowire (brown) with one side covered in aluminium (blue) placed on top of two plunger gates ($V_\text{lplg}$ and $V_\text{rplg}$) that tune the chemical potential in sections (green) of the proximitized InAs. A small region of the superconductor between these two segments is removed to create a Josephson junction, controlled by $V_\text{cut}$. A magnetic field, $B$, is applied along the wire axis.
}\vspace{-4mm}
\end{figure}

In this work, we present a high-magnetic-field-resilient nanowire-based transmon circuit. We demonstrate coherent qubit operation for in-plane magnetic fields up to 1~T. Further, we observe a field dependent periodic lobe structure in the qubit spectrum, attributable to interference effects as an integer number of flux quanta thread the nanowire cross section. Finally, we observe a rich spectrum of additional energy excitations as we transition into the first and second lobes of the qubit spectrum. We associate these excitations with Andreev states, visible due to their coupling to the qubit.

\begin{figure*}
\centering\vspace{-0mm}
\includegraphics[width=0.8\textwidth,angle=0]{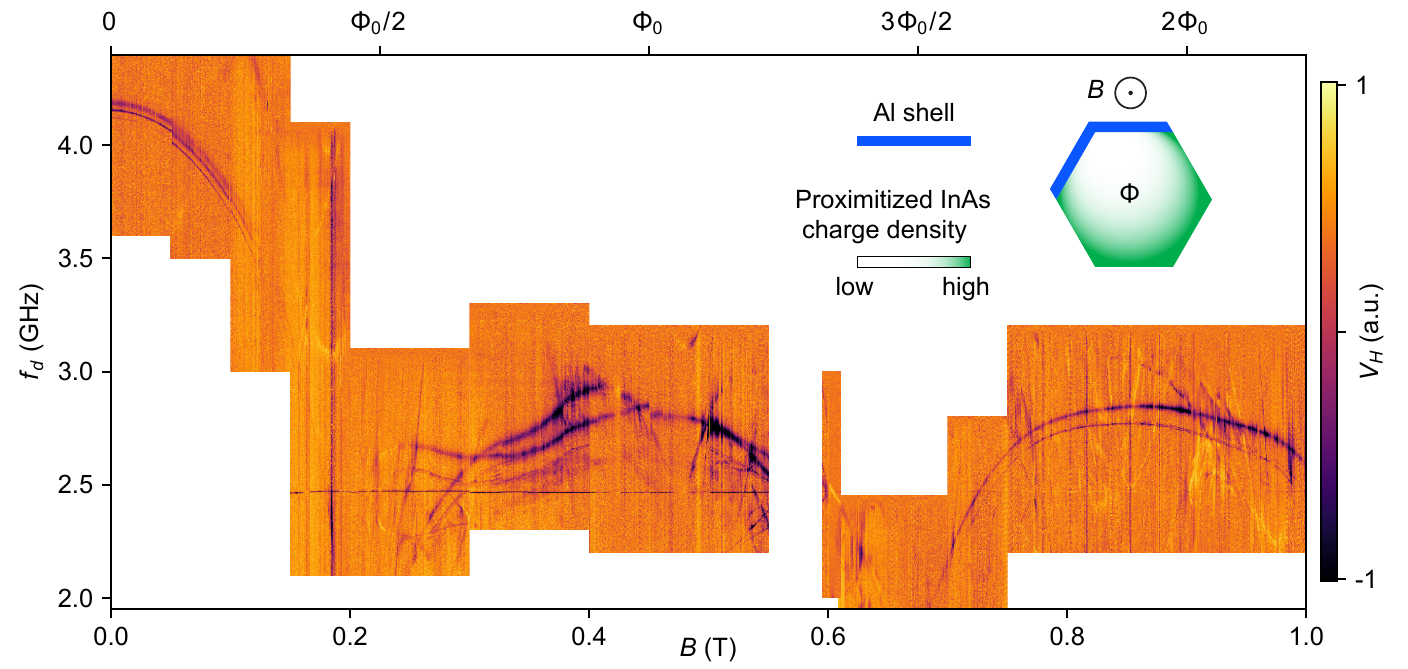}\vspace{-2mm}
\caption[Transmon in magnetic field]{\label{fig:TeslaScan}
Two-tone spectroscopy as a function of magnetic field $B$ at $V_\text{cut} = - 0.5$~V and $V_\text{lplg}=V_\text{rplg}= -4.2$~V. A variable drive tone, $f_d$, was applied and immediately followed by a readout tone at the cavity resonance frequency, allowing readout of the demodulated transmission voltage, $V_H$. The qubit drive power was adjusted between traces to account for varying lifetime and detuning from the readout cavity, which may cause changes in the background signal and linewidths of transitions. Data around $0.58$~T omitted due to too low applied signal power during two-tone spectroscopy. A line average is subtracted from data for each $B$. Inset: Sketch illustrating the cross section of a two-facet nanowire with the hypothesis that the electron density accumulates at the InAs surface, as illustrated by the color gradient (green/white). A superconducting ring is created by the superconducting Al shell (blue) and the proximitized InAs (green). For an axial magnetic field, $B$, a flux, $\Phi$, threads the nanowire cross section resulting in a periodic modulation of the qubit frequency. The top horizontal axis is constructed by inferring that a half-integer number of flux quanta, $\Phi_0(=h/2e)$, thread the nanowire at $B=225$~mT.
}
\vspace{-4mm}
\end{figure*}

\section{Magnetic field compatible gatemon device}

Figure~\ref{fig:SampleSchematics} shows the qubit device. A $20$~nm-thick NbTiN-film on a high resistivity silicon substrate was patterned by electron-beam-lithography and a chlorine-based dry-etch process to form the $\lambda/2$ readout resonator, qubit island, and bottom gate electrodes, see Fig.~\ref{fig:SampleSchematics}(a). We additionally patterned a high density of flux pinning holes, allowing high-field compatibility of the qubit capacitor and readout resonator~\cite{Kroll:2018aa,Kroll:2018sd}. NbTiN crossovers shorted ground planes together to prevent parasitic chip modes~\footnote{We found that using aluminum on-chip bond wire crossovers to connect ground planes caused a large amount of dissipation above the critical field of aluminum.}. 

A nanowire was placed on the NbTiN bottom gates using a micromanipulator. The nanowire has an InAs core, diameter $\sim 100$~nm, with two out of six facets covered by a $7$~nm thick epitaxially matched aluminum film~\cite{Krogstrup:2015en}. Prior to the initial NbTiN deposition, a local region of $5$~nm thick HfO$_2$ was deposited using atomic-layer-deposition techniques to ensure no leakage between the closely spaced gates through the silicon substrate~\footnote{Electrodes spaced $\sim 1~\mu$m apart on bare, high resistivity silicon will leak at $\sim \pm 10$~V at base temperature.}. A second thicker HfO$_2$ layer ($15$~nm) was deposited on top of the bottom gates as a gate dielectric. To form the Josephson junction, a small segment of the aluminum shell was removed by wet etching ($\sim100$ nm)~\cite{Larsen:2015cp}. 

To complete the gatemon qubit circuit, the nanowire was connected to the T-shaped qubit island, with simulated charging energy $E_C/h = 230$~MHz~\cite{comsol}, and to the surrounding ground plane, see Fig.~\ref{fig:SampleSchematics}(b). A light RF mill was used to remove the native oxide of InAs prior to depositing $\sim200$~nm NbTiN sputtered contacts. The qubit island was capacitively coupled to the $\lambda/2$ cavity with resonance frequency $f_r\sim 4.95$~GHz for readout and microwave control. Large plunger electrodes, $V_\text{lplg}$ and $V_\text{rplg}$, allowed for tuning of the chemical potential of the two proximitized nanowire segments on each side of the Josephson junction [green segments of Fig.~\ref{fig:SampleSchematics}(c)]. A third electrode, $V_\text{cut}$, located under the junction tuned the Josephson energy, $E_J$, and in turn the qubit frequency, $f_q$. On-chip $LC$-filters (not shown) on each gate electrode suppressed microwave dissipation through the capacitively coupled gates~\cite{Mi:2017aa}. A second qubit with no plunger gates was coupled to the same resonator (not shown). 

We present data from the qubit device shown in Fig.~\ref{fig:SampleSchematics}, which maintained coherence up to magnetic fields of $1$~T. For multiple similar devices we observed coherent operation up to $\sim 500$~mT. The sample was placed inside a CuBe enclosure filled with microwave absorbing Eccosorb foam to reduce stray microwave and infrared radiation. The enclosure was mounted inside a bottom-loading dilution refrigerator equipped with a 6-1-1 T 3-axis vector magnet and with a base temperature $<50$~mK (see Appendix~\ref{app:setup} for further details, including a schematic of the setup).

\section{Qubit measurements in large magnetic fields}

We investigated the qubit behavior by performing two-tone spectroscopy as a function of magnetic field, $B$, aligned along the nanowire axis. A varying drive tone at frequency $f_d$ was applied, followed by a readout tone for each $B$. During these measurements, the cavity resonance was first measured for each $B$ in order to correct for any changes in the readout frequency. Out-of-plane magnetic fields on the order of $10~\mu$T modified the resonance frequency of the cavity, however we did not observe any degradation in the resonator Q factor as the total magnetic field was varied. While changing B, intermittent corrections to the magnetic field alignment were also applied to minimize the out-of-plane magnetic field component (see Appendix~\ref{app:readout} for details). 

Figure~\ref{fig:TeslaScan} shows the qubit spectrum as a function of $B$ up to 1~T. The qubit spectrum exhibits a lobe structure with three lobes separated by minima at $B \sim 0.225$~T and $B \sim 0.675$~T and a reduced maximum qubit frequency in higher lobes. These minima may occur due to a suppression of the induced superconducting gap, $\Delta^{*}$, in the leads of the junction due to interference effects~\cite{gul_2014}. Depending on gate voltage the charge density in the semiconductor nanowire leads may be confined to the surface, see Fig.~\ref{fig:TeslaScan} inset. As analysed by Winkler \textit{et al.}~\cite{Winkler:2018ac}, a segment of this cross-sectional geometry effectively forms a superconducting ring interrupted by a semiconductor Josephson junction with the superconducting gap modulated by the periodic flux-biased phase difference [Fig.~\ref{fig:TeslaScan}, inset]. For the case of half a flux quantum threading the nanowire at $B=0.225$~T, the applied flux $\Phi$ in units of flux quanta $\Phi_0$ is shown along the top horizontal axis of Fig.~\ref{fig:TeslaScan}. From this period, we estimate the effective diameter of the interference loop to be $d_\text{eff} = \sqrt{2\Phi_0/\pi B(\Phi=\Phi_0/2)} = 76~$nm. As the charge accumulation layer will have a finite thickness one expects a slightly smaller effective diameter compared to that of the nanowire ($\sim 100~$nm)~\cite{Antipov_2018}. Simulations of realistic wire geometries~\cite{Winkler:2018ac} also predict a reduced maximum superconducting gap in higher lobes due to inhomogeneity in the effective diameter. This is consistent with our measured data where the qubit frequency, $f_q$, is expected to scale with $ \sqrt{\Delta^{*}}$. Similar oscillations with magnetic field have also been observed for nanowires in transport experiments~\cite{zou_2017} and were attributed to interference effects in the junction itself, which may also play a significant role here. We note that the field dependence is strongly influenced by the nanowire charge distribution and the oscillations observed here were for a particular range of plunger-gate values~\cite{Winkler:2018ac, danilenko}. Periodic oscillations in qubit frequency have also been observed for gatemons with nanowire junctions where the Al shell fully enclosed the leads, which were interpreted as the Little-Parks effect~\cite{sabonis2020}.

\begin{figure}
\centering
\includegraphics[width=1\columnwidth]{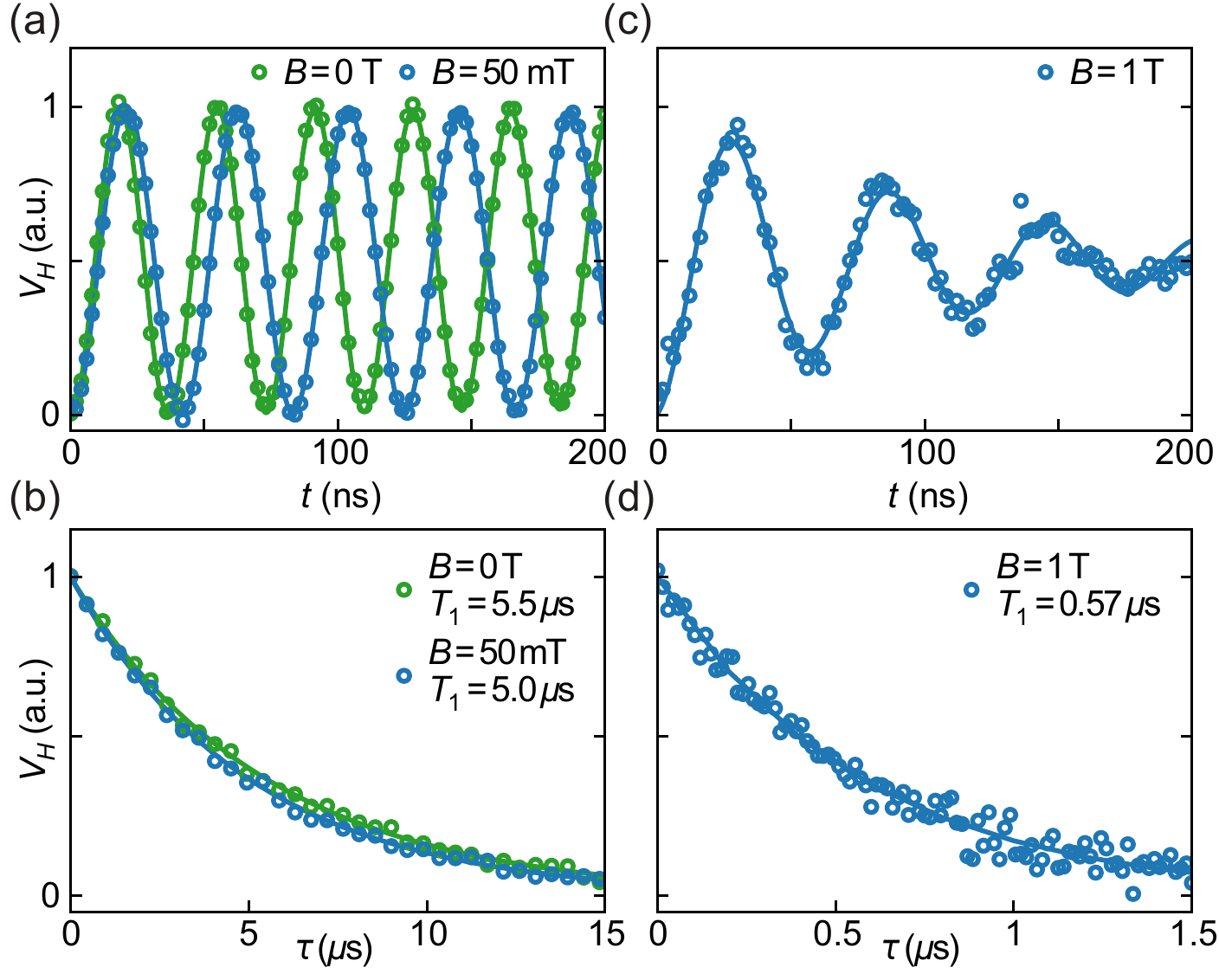}\vspace{-2mm}
\caption[Low field coherence]{\label{fig:Time_domain} Time domain measurements as a function of $B$ at $V_\text{cut} = - 0.5$~V and $V_\text{lplg}=V_\text{rplg}= -4.2$~V.
(a) Rabi oscillations of the gatemon at $B=0$ and $B=50$~mT. We measured the demodulated transmission, $V_H$, as a function of drive duration, $t$, applied at the qubit frequency. Fits are exponentially damped sinusoids. Data are normalized to the extracted fit parameters. (b) $T_1$-lifetime measurement at $B=0$ and $B=50$~mT. We measured $V_H$ as a function of delay time, $\tau$, between the drive and readout tones. To excite the qubit, we applied a $\pi$-pulse calibrated from (a) at $f_q=4.2$~GHz (green) and $f_q=4.1$~GHz (blue). The data are fitted to a decaying exponent to extract $T_1$. (c) [(d)] same as (a) [(b)] at $B=1$~T with $f_q=2.4$~GHz.}\vspace{-4mm}
\end{figure}

We next consider the qubit behavior in each of the three lobes. In the zeroth lobe measured from $B\sim 0$ to $\sim 150$~mT, the qubit behaves indistinguishably from a standard gatemon device. Due to the high drive power, multi-photon transitions are observed, exciting higher energy states of the qubit. Around $150$~mT the system became unmeasurable due to the second qubit on the chip anti-crossing with the readout resonator, see Fig.~\ref{fig:FvB_res}. Figures~\ref{fig:Time_domain}(a) and~\ref{fig:Time_domain}(b) show Rabi oscillations and lifetime decay at $B=0$ and $B=50$~mT. At $B=0$ we observe lifetimes of $\sim 5.5~\mu$s similar to previous gatemon devices with a single junction gate, indicating that the additional plunger gates and dielectric layers do not compromise qubit performance. The measurements show almost no difference between $B = 0$ and $B = 50$~mT demonstrating excellent resilience to parallel magnetic fields consistent with other recent studies of gatemon qubits~\cite{Luthi:2018aa}. Furthermore, as the field was not perfectly aligned, these data indicate that small out-of-plane magnetic fields ($\sim10~\mu$T) do not degrade qubit quality. This suggests that our qubit design mitigates the need for extensive magnetic shielding, as typically required for superconducting qubit devices.

Moving to the first lobe between $B\sim 250$~mT and $B\sim 650$~mT, two main resonances appear (Fig.~\ref{fig:TeslaScan}). Both states behave as weakly anharmonic oscillator modes with a broad single-photon transition frequency and a sharper two-photon transition separated by $\sim 100$~MHz. While the presence of two anharmonic states is consistent with a large Majorana coupling across the junction mediated by two overlapping zero modes~\citep{Ginossar:2014aa}, it is unlikely that the splitting is due to Majorana physics as the topological phase is typically expected to occur at higher magnetic fields for InAs-based wires. Rather, the splitting might be connected to low-energy Andreev states interacting with the qubit mode, as indicated by several transitions dispersing strongly with magnetic field throughout the first lobe. In this regime, it was not possible to probe the qubit states using time domain measurements due to very low lifetimes.

In the second lobe above $B\sim 650$~mT, a single qubit resonance revives and is clearly visible all the way up to $B = 1$~T. The two-photon $0\to 2$ transition is also observed below the qubit transition. Similar to the first lobe, additional resonances strongly dispersing in magnetic field are also observed in the second lobe. Figures~\ref{fig:Time_domain}(c) and~\ref{fig:Time_domain}(d) show coherent Rabi oscillations of a superconducting transmon qubit at $B=1$~T with lifetime $T_1 = 0.57~\mu$s. We speculate that the decrease in $T_1$ a $B=1$~T is due to a reduction in $\Delta^{*}$ compared to at $B=0$, resulting in an increase in quasiparticle poisoning rates~\cite{catelani_2011, uilhoorn}.

\begin{figure}
\centering
\includegraphics[width=1\columnwidth]{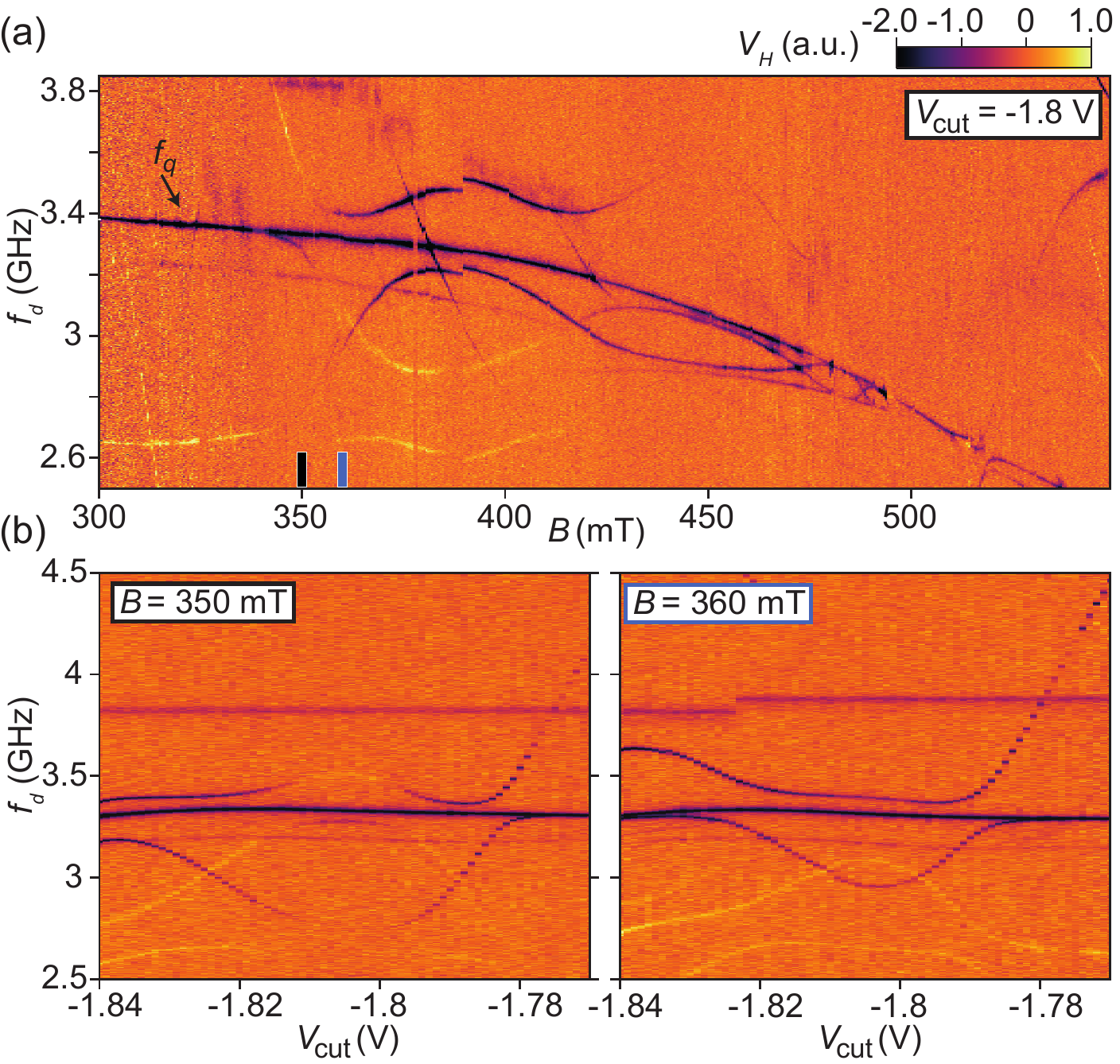}\vspace{-4mm}
\caption[Junction states vs magnetic field]{\label{fig:FvsB}
Junction states as a function magnetic field and gate. (a) Two-tone spectroscopy for varying $f_d$ and $B$ reveals oscillating behavior of junction states at gate voltage $V_\text{cut} = - 1.8$~V and $V_\text{lplg}=V_\text{rplg}= -2.0$~V. We observe an uninterrupted qubit transition frequency $f_q$ (arrow) decaying as $B$ increases with multiple new transitions emerging and exhibiting multiple avoided crossings with the qubit transition. (b) Two-tone spectroscopy as a function of $f_d$ and $V_\text{cut}$ at $B=350$~mT [left, black rectangle in (a)], and $360$~mT [right, blue rectangle in (a)]. Again, a clear qubit transition is visible, weakly dependent on $V_\text{cut}$ with two strongly dispersing transitions coupling to the qubit. The gate-independent transition at $f_d\sim3.8$~GHz is the qubit transition of the second qubit. A line average is subtracted from each column in all panels.}\vspace{-4mm}
\end{figure}

\section{Anomalous junction states}
\vspace{-2mm}
To investigate the anomalous qubit-resonance splittings in the first lobe, we focus on a voltage regime where sharp additional transitions and avoided crossings in the qubit transition are observed, as shown in Fig.~\ref{fig:FvsB}(a). A clear, uninterrupted qubit transition frequency $f_q$ is slowly reduced from $f_q\sim 3.4$ to $2.8$~GHz, as $B$ is increased from 300 to 500~mT [arrow in Fig.~\ref{fig:FvsB}(a)]. Additionally, around the qubit transition, several new resonances appear for $B>350$~mT, oscillating with magnetic field. When these oscillating state transitions are on resonance with the qubit, we observe avoided crossings, indicating strong coupling to the qubit. We associate these resonances with low-energy Andreev bound states that couple to the resonator via the qubit, in agreement with recent numerical simulations of similar nanowire structures~\cite{keselman2019spectral}. We speculate that the coexistence of the coupled and uncoupled spectra,  as seen emerging at $B\sim350$~mT and $f_q\sim3.4$~GHz in Fig.~\ref{fig:FvsB}(a), can be explained by a fluctuating parity of the Andreev states~\cite{Janvier:2015uf,vanWoerkom:2017ek,Hays:2017ud}. For instance, in the even parity state, additional transition frequencies might be observed due to the hybridization of the qubit and Andreev state transitions. However, in the odd parity state (or vice versa~\cite{Tosi_2019,Hays2020}), a single qubit resonance might only be observed as the Andreev state transitions no longer hybridize with the qubit mode. The observed spectrum is then the average of these different configurations with parity switching occurring faster than the measurement time (typically $>10$ s for each vertical trace). 

To further probe the spectrum, we swept $V_\text{cut}$ at fixed $B$, see Fig.~\ref{fig:FvsB}(b). Here, the qubit transition is weakly dispersing around $f_q\sim3.4$~GHz. Two strongly gate-dependent transitions oscillate around $f_d\sim3$~GHz with both transitions giving rise to avoided crossings with the qubit transition. In addition, a second weakly gate dependent state at $f_d\sim2.5$~GHz exhibits avoided crossings with the two oscillating transitions. The strong dispersion of the transitions with $V_\text{cut}$ is consistent with Andreev states that are localized in the junction and therefore expected to be strongly dependent on the electrostatics of the junction.

\section{Conclusions}

We have presented a magnetic-field resilient gatemon circuit with excellent relaxation times of $5~\mu$s at moderate magnetic fields, $\sim 50$~mT. The qubit retains coherence up to magnetic fields of $1$~T with a lifetime $T_1 \sim 0.6~\mu$s, demonstrating compatibility of our gatemon circuit design with magnetic fields typically needed for Majorana zero modes. Future work could integrate additional gates to allow greater control of the charge carrier distribution along the nanowire or use a SQUID-like geometry to allow control of the superconducting phase across the Josephson junction. Combining the microwave spectroscopy techniques with dc transport measurements~\cite{lead} may also provide further insights into the underlying origin of observed features.

\begin{acknowledgments}

We thank Arno Bargerbos, Bernard van Heck, Angela Kou, Leo Kouwenhoven, and Gijs de Lange for valuable discussions. Research was supported by Microsoft, the Danish National Research Foundation, and the European Research Council under grant HEMs-DAM No.716655.
\end{acknowledgments}

\appendix

\section{Readout frequency corrections}
\label{app:readout}

\begin{figure}[h]
\centering
\includegraphics[width=\columnwidth]{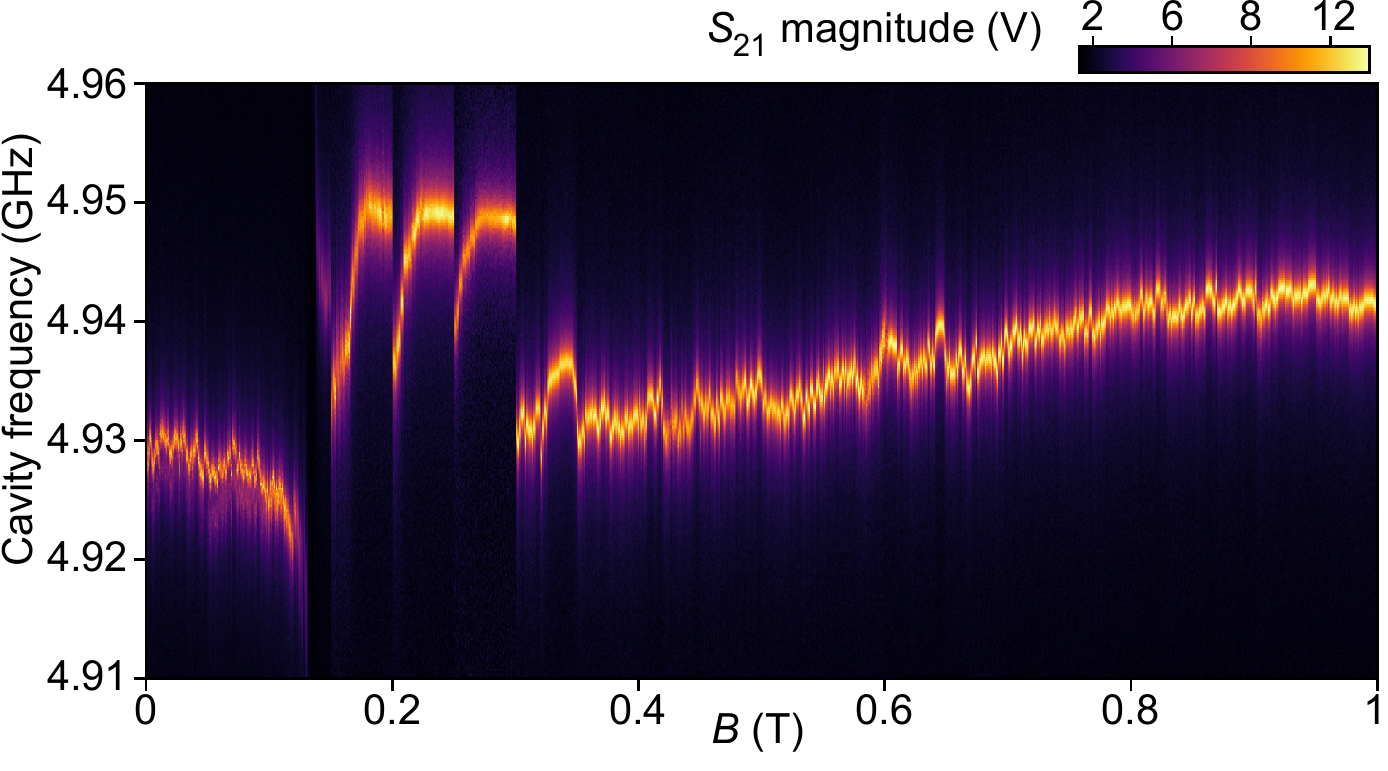}\vspace{-2mm}
\caption[resonator]{\label{fig:FvB_res} Transmission voltage, $S_{21}$, as a function of cavity drive frequency and $B$ showing the field modulation of the resonance frequency of the $\lambda/2$ cavity used for readout in Fig.~\ref{fig:TeslaScan}. Due to the large fluctuations, the readout frequency was adjusted each time the magnetic field was varied. Large jumps around $B=0.2$~T are due to corrections of the out-of-plane magnetic field $\sim0.1~$mT (not shown). At $B\sim150$~mT the avoided crossing between the resonator and the second qubit is observed.}
\vspace{-4mm}
\end{figure}

When applying an in-plane magnetic field, $B$, the readout resonator frequency, $f_r$, was modulated due to changes in the kinetic inductance of the NbTiN film. We therefore corrected the readout frequency before each two-tone spectroscopy measurement by measuring the transmission voltage, $S_{21}$, as a function of frequency with a vector network analyzer, see Fig.~\ref{fig:FvB_res}. Following each measurement, we fitted $S_{21}$ to a skewed Lorentzian to determine the readout frequency. These measurements were interleaved with the two-tone spectroscopy measurements shown in Fig.~\ref{fig:TeslaScan}. We observe a slight degradation in $f_r$ until the avoided crossing with the second qubit at $B\sim150$~mT is observed. The large jumps in $f_r$ around $B\sim200$~mT are due to corrections to the out-of-plane field components carried out in between measurements, after observing a steady decrease in $f_r$ when sweeping down from $B=1$~T. No significant degradation in the peak width is observed, highlighting the magnetic field compatibility of the readout resonators.

\section{Experimental setup}
\label{app:setup}

\begin{figure}[!h]
\centering
\includegraphics[width=\columnwidth]{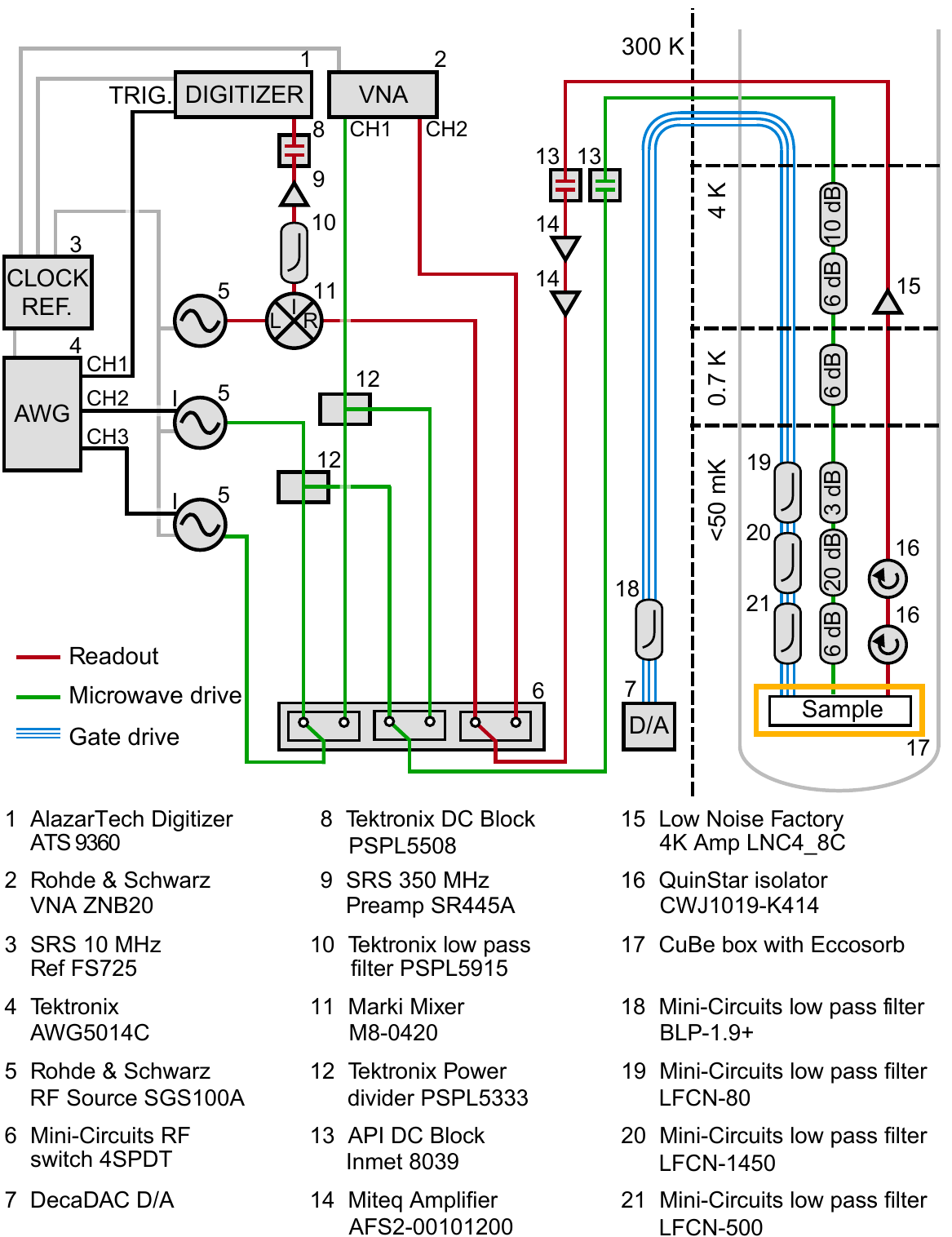}\vspace{-2mm}
\caption[setup]{\label{fig:setup}
Schematic of the experimental setup used for the experiments presented. The readout resonator was driven either by the VNA or an AWG-modulated RF source (green lines). The output signal (red lines) was amplified and read out either with the VNA or down converted by mixing with a reference signal. All microwave equipment was synchronized with a 10~MHz clock reference. Three DC lines (blue) were connected to the three gates, $V_\text{lplg}$, $V_\text{rplg}$, and $V_\text{cut}$, to tune the nanowire chemical potential and junction, respectively.
}
\end{figure}

Figure~\ref{fig:setup} shows the experimental setup used for the measurements presented in the paper. The readout resonance frequency was determined by transmission measurements with a vector network analyzer (VNA). Two-tone spectroscopy and time domain measurements were acquired with a heterodyne demodulation readout circuit. With this circuit we measured the transmission of a pulse modulated RF signal. We amplified the transmitted signal at 4~K and further at room temperature and then mixed down with a reference signal before sampling and digital down conversion. The demodulation circuit and VNA were connected to an RF switch matrix to allow switching between the two measurement configurations. Experiments were carried out in a dilution refrigerator with a 6-1-1 T vector magnet.

%

\end{document}